\documentclass[12pt]{iopart}
\newcommand \be{\begin{eqnarray}}
\newcommand \ee{\end{eqnarray}}
\newcommand \ba{\begin{align}}

\newcommand {\V}[1]{{\bf #1}}

\begin{document}

\title{Chiral anomaly in Weyl systems: no violation of classical conservation laws}

\author{K. Morawetz$^{1,2,3}$}

\address{$^1$M\"unster University of Applied Sciences,
Stegerwaldstrasse 39, 48565 Steinfurt, Germany\\
$^2$International Institute of Physics- UFRN,
Campus Universit\'ario Lagoa nova,
59078-970 Natal, Brazil\\
$^{3}$ Max-Planck-Institute for the Physics of Complex Systems, 01187 Dresden, Germany}
\ead{morawetz@fh-muenster.de}
\vspace{10pt}
\begin{indented}
\item[]July 2018
\end{indented}

\begin{abstract}
The anomalous term $\sim\V E\V B$ in the balance of the chiral density can be rewritten as quantum current in the classical balance of density. Therefore it does not violate classical conservation laws as it is claimed to be caused by quantum fluctuations.
\end{abstract}

%
%
%
%
%

Relativistic Fermions with zero mass and consequently linear dispersion have a definite chirality by parallel or anti-parallel spin and motion directions \cite{W29}.   
In condensed matter physics an excitation of chiral mass-less Dirac particles with linear dispersion has been discovered in a class of Weyl semi-metals like $TaAs$ \cite{Xu613,Lu622,Lv15,Hu15,We15} after being predicted in \cite{Wa11}. The chirality is measured here by the photocurrent in response to circularly polarized mid-infrared light \cite{Ma17}. 

From field theory one knows that the transition amplitude is changed by chiral rotational transformation which means that the chiral current is not conserved.  
Let us shortly remind the main phenomenological idea following the heuristic derivation of \cite{Fuk08}. A parallel electric and magnetic field changes the chirality. The Fermi momentum of the right-handed Fermions increases in the electric field
\be
p_F=eEt
\ee
with opposite direction for left-handed ones. The density of left and right-handed Fermions is the product of longitudinal phase-space density, $dN_R/dz=p_F/2\pi\hbar$, and the density of Landau levels in traverse direction, $d^2N_R/dx dy=eB/2\pi\hbar$, such that the rate of chirality $N_5=N_R-N_L$ is
\be
{d n_5\over d t}={d^4 N_5\over d t d^3x}=2 {\dot p_f\over 2 \pi \hbar}{e B\over 2 \pi \hbar}={e^2\over 2 \pi^2\hbar^2} \V E \cdot \V B.
\label{rate}
\ee
Therefore the term $\V E\V B$ is considered as the origin of non-conservation of chiral charge.

The $\V E\V B$ term in the chiral density balance (\ref{rate}) also is the basis of the experimental interpretation \cite{Hu15,Xu17,gooth17} of having observed chiral anomaly and breaking of conservation laws like mixed axial-gravitational anomaly. Since the path from this symmetry-violating assumptions to the final `non-conservation` form is well worked out \cite{ZB12a}, it leads to the impression as if the opposite conclusion would be unique as well and it would be a unique signal of violation of conservation laws. However, this is not a one-to-one correspondence. One cannot conclude backwards from the observed term (\ref{rate}) to a symmetry-breaking field-theoretical assumption. In fact there is another path to obtain the same term (\ref{rate}) from a conserving theory without the described field-theoretical assumptions \cite{M18}.

In field theory the origin of this seemingly charge-nonconserving term is the chiral anomaly based on anomalous Ward identities which quantum vector or axial vector field obey. Only exclusively one of them can be made normal \cite{Ba69}.
Frequently the  charge anomaly is introduced  in the relativistic Lagrangian \cite{ZB12,CPWW13,Bu15,Bu16,JHK17} as axial chemical potential term or effective theta angle \cite{Fuk08}
\be
{e^2\over 16 \pi^2\hbar^2 c}(\V b\V r-b_0 t)\varepsilon^{\mu\nu\alpha\beta}F_{\mu\nu}F_{\alpha\beta}
\label{axion}
\ee
with chiral gauge fields 
$(b_0,\V b)$ 
also called axion fields \cite{GT13}. The electrodynamics assuming explicitly such a chiral breaking term has been treated in \cite{QCH17}. These terms have been suggested by anomalous terms coming from triangular anomaly \cite{J12,BKY14,L16} known as Adler-Jackew-Bell anomaly \cite{Ad69,BJ69,NN83}.  Due to the axial non-conservation it is called also mixed axial-gravitational anomaly and claimed to violate Lorentz symmetry \cite{ZB12,JHK17,Xu17,gooth17}. 
In \cite{DelCima2017} it has been shown recently that a proper subtraction scheme of the infrared divergences shows that such extra terms do not appear. Therefore the claim of Lorentz-symmetry violation anomaly is not well founded theoretically. 
In fact the Lorentz-invariant chiral kinetic theory can be derived from the quantum kinetic approach \cite{M18,GLPWW12,CPWW13,MT14,GPW17,HPY17,HSJLZ18}.

Here we want to point out that the anomalous term (\ref{rate}) can be rewritten itself as a quantum current fitting perfectly the classical balance equation
\be
\partial_t n_5+{\V \nabla}\cdot (\V j+\V j_{\rm anom})=0.
\label{3}
\ee

The algebra is astonishingly simple. We consider the magnetic field given by a vector potential $\V B={\V \nabla}\times \V A$ and the electric field by an additional scalar field $\V E=-{\V \nabla} \Phi-\dot {\V A}$ for any gauge. Since ${\V \nabla} (\Phi {\V \nabla}\times \V A)={\V \nabla} \Phi \cdot ({\V \nabla}\times \V A)$ we have
\be
\V B\cdot \V E=-(\V \nabla\times \V A)\cdot (\dot {\V A}+\V \nabla\Phi)=-({\V \nabla}\times \V A)\cdot \dot {\V A}-{\V \nabla} (\Phi {\V \nabla}\times \V A).
\label{2}
\ee
The first term can be expressed as a divergence too in the following way. The standard relation 
allows us to write
\be
({\V \nabla}\times  \V A) \cdot \dot {\V A}
&=&-{\V \nabla} \cdot (\dot {\V A}\times \V A)+\V A \cdot ({\V \nabla} \times \dot {\V A})
\nonumber\\
&=&-{\V \nabla} \cdot (\dot {\V A}\times \V A)-\dot{\V A }\cdot ({\V \nabla}\times {\V A})
\label{1}
\ee
where we have used the time derivative of $d/dt[\V A\cdot ({\V \nabla} \times \V A)]=0$ to obtain the second line. From (\ref{1}) we see that the second term of the last line yields together with the left side just one half of the first term of the right side,
\be
({\V \nabla}\times  \V A) \cdot \dot {\V A}
&=&-\frac 1 2{\V \nabla} \cdot (\dot {\V A}\times \V A).
\ee 
Using this in (\ref{2}) we can write the anomalous term (\ref{rate}) into a form of a divergence
\be
{e^2\over 2 \pi^2\hbar^2}\V B\cdot \V E=-{\V \nabla} \cdot \V j_{\rm anom}
\ee
with an anomalous current
\be
\V j_{\rm anom}={e^2\over 2 \pi^2\hbar^2}\left (\frac 1 2 \dot {\V A}\times \V A-\Phi {\V \nabla}\times \V A \right ).
\label{j}
\ee
This current together with (\ref{3}) is the result of this letter.
Since it is an exact rewriting, we cannot interpret the $\V E\V B$ term of (\ref{rate}) observed in solid state physics as a signal of violation of classical conservation laws since it fits perfectly the classical balance (\ref{3}). One observes merely an extra current by anomalous transport \cite{M15}. 

It has to be noted that this quantum current (\ref{j}) is not gauge invariant and cannot be obtained by a variation of a Lagrangian. This underlines the anomaly, but does not violate classical conservation laws. The same observation one can find in literature on general covariant forms. One can define modified chiral currents which can be written in a conserved form, for a recent discussion see \cite{FB00}.  This is connected with the fact that one can rewrite the anomaly in a consistent way \cite{WZ71} which means as a variation of a functional with respect to a gauge field. This considers the effective action on the edge and not in the bulk. The variation of the bulk action contributes to the edge current and is described by Bardeen polynomials such that a covariant form can be achieved \cite{BZ84}. 
In \cite{L14} it was pointed out that the divergence of covariant currents is uniquely defined, while the divergence of consistent currents depends on
specific regularization schemes which freedom allows the definition of an
exactly conserved electric current. 

All these treatments should lead to the physical observation that charges separate at the edges perpendicular to the magnetic field
even when there is no bulk current according to kinetic theory.
This is in fact illustrated by the successful application of the chiral anomaly to describe universality of conductance \cite{ACF98} and anomalous transport including Hall effects \cite{La16}. Here we have seen that this additional current (\ref{j}) is an exact rewriting of the chiral anomaly and fits perfectly the classical balance equation for density (\ref{3}) and therefore the conservation law.

Finally I would like to thank J\"urg Fr\"ohlich and Tobias Meng for valuable comments and outline of important literature. 
\section*{References}                
\bibliography{entropy,bose,kmsr,kmsr1,kmsr2,kmsr3,kmsr4,kmsr5,kmsr6,kmsr7,delay2,spin,spin1,refer,delay3,gdr,chaos,sem3,sem1,sem2,short,cauchy,genn,paradox,deform,shuttling,blase,spinhall,spincurrent,tdgl,pattern,zitter,graphene,quench,msc_nodouble,iso,march,weyl,anomal}
\bibliographystyle{Science}

\end{document}